\documentclass[
superscriptaddress
]{revtex4-1}

\usepackage{graphicx}
\usepackage{dcolumn}
\usepackage{bm}

\begin{document}

\newcommand{\ket}[1] {
  | #1 \rangle}
\title{All-mechanical coherence protection and fast control of a spin qubit}
 \author{Eliza Cornell}\thanks{These authors contributed equally to this work.}\affiliation{John A. Paulson School for Engineering and Applied Sciences, Harvard University, Cambridge, Massachusetts, USA.}

\author{Zhujing Xu}\thanks{These authors contributed equally to this work.}\affiliation{John A. Paulson School for Engineering and Applied Sciences, Harvard University, Cambridge, Massachusetts, USA.}

\author{Zhaoyou Wang}
\affiliation{Pritzker School of Molecular Engineering, University of Chicago,
Chicago, Illinois, USA.}

\author{Hana K. Warner}
\affiliation{John A. Paulson School for Engineering and Applied Sciences, Harvard University, Cambridge, Massachusetts, USA.}

\author{Eliana Mann}
\affiliation{John A. Paulson School for Engineering and Applied Sciences, Harvard University, Cambridge, Massachusetts, USA.}

\author{Michael Haas}
\affiliation{John A. Paulson School for Engineering and Applied Sciences, Harvard University, Cambridge, Massachusetts, USA.}
\affiliation{IonQ, College Park, Maryland, USA.}

\author{Smarak Maity}
\affiliation{John A. Paulson School for Engineering and Applied Sciences, Harvard University, Cambridge, Massachusetts, USA.}
\affiliation{ Q-CTRL, Los Angeles, California, USA.}

\author{Graham Joe}
\affiliation{John A. Paulson School for Engineering and Applied Sciences, Harvard University, Cambridge, Massachusetts, USA.}
\affiliation{Nubis Communications, New Providence, New Jersey, USA.}

\author{Liang Jiang}
\affiliation{Pritzker School of Molecular Engineering, University of Chicago,
Chicago, Illinois, USA.}

\author{Peter Rabl}
\affiliation{Walther-Meißner-Institut, Bayerische Akademie der Wissenschaften, Garching, Germany.}
\affiliation{TUM School of Natural Sciences, Technische Universit{\"a}t M{\"u}nchen, Garching, Germany.}
\affiliation{Munich Center for Quantum Science and Technology (MCQST), Munich, Germany.}

\author{Benjamin Pingault}
\affiliation{Argonne National Laboratory, Lemont, Illinois, USA.}

\author{Marko Lončar}
\affiliation{John A. Paulson School for Engineering and Applied Sciences, Harvard University, Cambridge, Massachusetts, USA.}

\date{\today}

\begin{abstract}
In a phononic quantum network, quantum information is stored and processed within stationary nodes defined by solid-state spins, and the information is routed between nodes by phonons\cite{habraken_continuous_2012,lemonde_phonon_2018}. The phonon holds distinct advantages over its electromagnetic counterpart the photon, including smaller device footprints, reduced crosstalk, long coherence times at low temperatures, and strong interactions with both solid-state spins and electromagnetic waves. Enhanced interactions between a phononic cavity and a stationary qubit have been demonstrated in multiple platforms including superconducting qubits\cite{manenti_circuit_2017,chu_quantum_2017,arrangoiz-arriola_resolving_2019,wollack_quantum_2022,lee_strong_2023}, spins in silicon carbide\cite{whiteley_spinphonon_2019,dietz_spin-acoustic_2023} and spins in diamond\cite{shandilya_optomechanical_2021,joe_observation_2025}. However, an outstanding issue is the compatibility between the spin's coupling to the resonant phononic cavity and the simultaneous use of pulse sequences to extend the coherence time of the spin by suppressing the low-frequency environmental noise. Here we demonstrate all-mechanical coherence protection of a solid-state spin qubit, where optical initialization, quantum operations, and readout are performed in a dressed basis that is highly immune to low-frequency noise and compatible with a phononic cavities. We additionally show record-high Rabi frequencies reaching 800 MHz, which allows for ultrafast quantum control. Our results establish a first step for high-fidelity, phonon-mediated quantum gates\cite{arrazola_toward_2024} and represent a crucial advance toward robust on-chip quantum phononic networks.
\end{abstract}

\maketitle

\section{\label{sec:intro}Introduction}
The phonon holds great interest as a quantum information carrier in on-chip quantum communications, in particular for its ability to efficiently interface between distinct quantum systems \cite{schuetz_universal_2015,rabl_quantum_2010}. Phononic quantum networks have been proposed in which quantum states are encoded in long-lived electronic spins and can then be converted into propagating phonon wave packets and reabsorbed efficiently by distant spin memories\cite{habraken_continuous_2012,lemonde_phonon_2018}. 

Realizing an efficient spin-phonon interface requires a strong coupling between an individual spin and a local phonon mode  \cite{lemonde_phonon_2018,schuetz_universal_2015,rabl_quantum_2010}, as well as control techniques to protect the spin from the ubiquitous magnetic field noise in solid-state environments\cite{sukachev_silicon-vacancy_2017}. Dynamical decoupling pulses are usually applied to achieve the latter task by periodically flipping the spin state and thereby refocusing the phase. However, a major challenge arises when the spin is embedded within a resonant phononic cavity, in which case the pulsed decoupling can conflict with the continuous spin-phonon coupling of interest or require very strong magnetic driving fields that are difficult to implement in a cryogenic environment.

An alternative scheme to mitigate the decoherence is to use a continuous-wave dynamical decoupling protocol, in which a coherent field is continuously applied to the spin and the quantum information is encoded in the dressed states defined by the interaction of the driving field with the spins\cite{arrazola_toward_2024}. This continuous dressing field decouples the spin from low-frequency noise to extend its coherence time and has been demonstrated with NV centers in diamond\cite{xu_coherence-protected_2012,barfuss_strong_2015,kim_suppression_2025}, divacancy centers in SiC\cite{miao_universal_2020}, semiconductor donor spin qubits\cite{laucht_dressed_2017} and trapped ions\cite{timoney_quantum_2011}. At the same time this approach is fully compatible with the realization of a tunable spin-phonon interface in which the energy splitting between the dressed spin states is tuned in and out of the phononic cavity resonance. This is achieved through a fast and broadband control of the driving amplitude.

Among spins in solid-state materials, the silicon vacancy center (SiV) in diamond (Fig. \ref{fig1}(a)) stands out for its utility as a long-lived quantum memory in a scalable quantum network. SiVs have an extremely high strain susceptibility\cite{meesala_strain_2018}, which makes them a promising candidate for phononic quantum networks in particular, and they have been integrated into GHz mechanical cavities\cite{joe_observation_2025}. However, continuous dynamical decoupling of the SiV spin has remained unexplored. 

In this work, we demonstrate the first realization of all-mechanical continuous coherence protection of a single SiV spin. In our approach, an acoustic mode resonant with the bare spin states creates a set of dressed states with an extended coherence time by reducing the spin’s sensitivity to magnetic field noise and the surrounding nuclear spin bath. The field that is used to probe the dressed states is also mechanical, validating that these coherent states can be efficiently coupled to phononic modes. The dressed states can be directly optically initialized and read out, which simplifies the experiment and eliminates the need for more complicated state preparation sequences or adiabatic state transfer\cite{miao_universal_2020,abdurakhimov_driven-state_2020,laucht_dressed_2017}. We observe a threefold extension in coherence time via the continuous decoupling scheme. This technique is fully compatible with future phononic networks and is a first step for high-fidelity, phonon-mediated quantum gates \cite{arrazola_toward_2024}. We also demonstrate a record-high Rabi frequency of 800 MHz at cryogenic temperatures, using our platform's efficient conversion from microwave to strain field to spin drive. This can enable ultrafast quantum control, which is relevant for future quantum networking applications of SiV spins.

\section{Results}\label{results}

\begin{figure}
\centering
\includegraphics[width=0.7\textwidth]{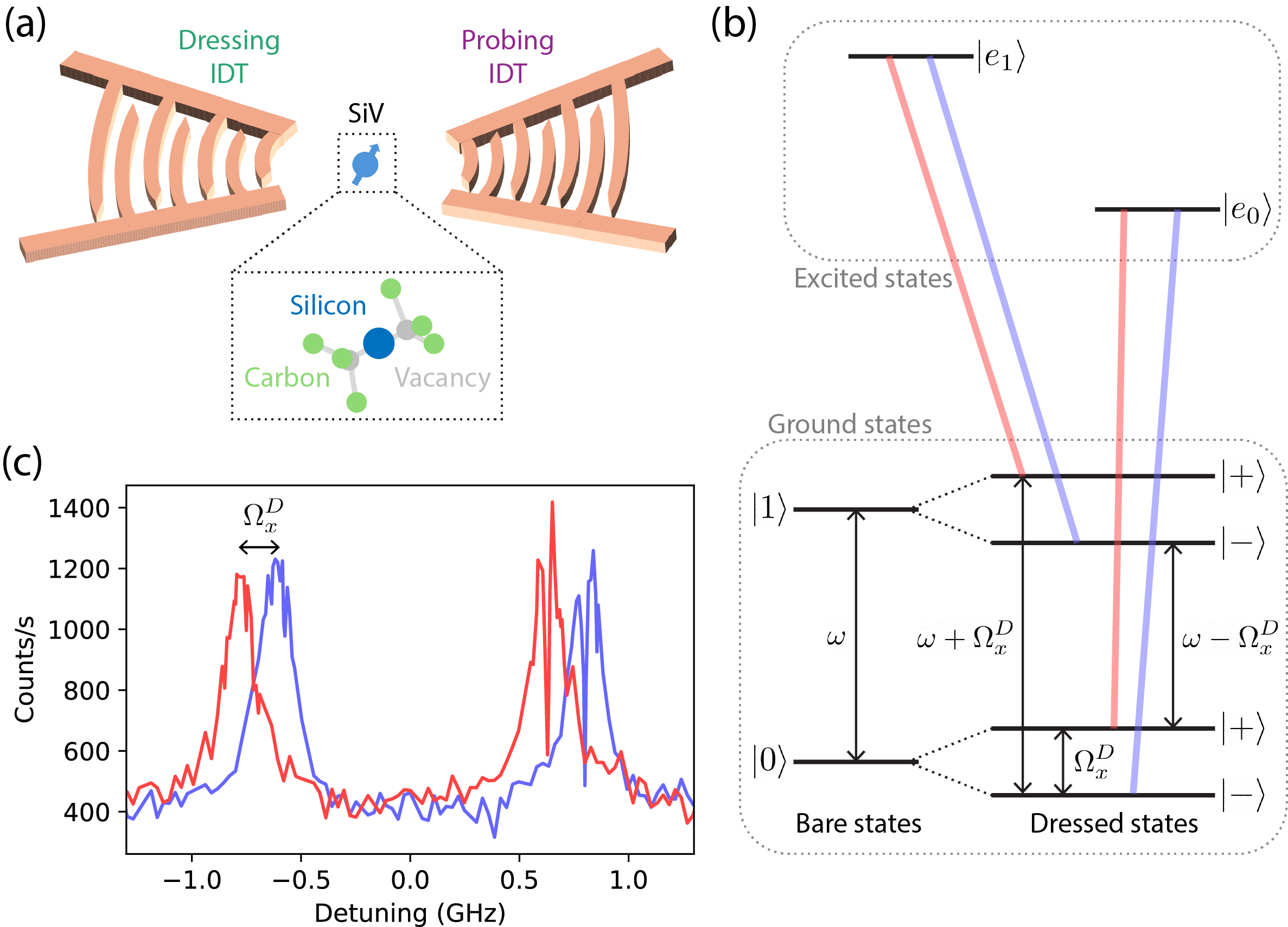}
\caption{Dressed states of a single SiV spin in diamond. \textbf{(a)} A microwave signal is applied independently to each of the interdigital transducers (IDTs) which generate acoustic waves. The resulting strain in the diamond can coherently control a single SiV spin near the focus of the IDTs. \textbf{(b)} Electronic level structure of a single SiV spin. The splitting $\omega$ between the bare spin states $\ket{0}$ and $\ket{1}$ can be Zeeman-tuned by an external magnetic field. A continuous-wave acoustic field with Rabi frequency $\Omega_x^D$ resonantly drives the spin and forms the dressed states $\ket{+}$ and $\ket{-}$, with energy levels offset by  $\pm\Omega_x^D/2$ from the bare states. Because of the time-dependence inherent in their definition, the dressed states do not have well-defined energies in the lab frame, so they can be resonantly driven at any of three different frequencies, $\Omega_x^D$ or $\omega \pm \Omega_x^D$. The dressed states have allowed optical transitions to the $\ket{e_0}$ and $\ket{e_1}$ excited states, which are not dressed because their splitting is detuned from $\omega$ by $>$1 GHz.  \textbf{(c)} A partial optical spectrum of the dressed states with $\Omega_x^D\sim$150MHz, showing the transitions indicated in (b) and taken via photoluminescence excitation by scanning the frequency of a resonant initialization laser. The red (blue) spectrum corresponds to the $\ket{+}$ ($\ket{-}$) states, and the frequency detuning is relative to a central optical wavelength of 737.0610nm. More information on the optical spectra measurement can be found in supplementary section \ref{ssec:optical}.
}\label{fig1}
\end{figure}

The strain fields used for quantum control of a single SiV are generated piezoelectrically, in a sputtered aluminum nitride layer overlaying a bulk diamond chip. Metal interdigital transducers (IDTs) are patterned on top, which transduce applied microwave fields into strain fields and generate propagating surface acoustic wave (SAW) modes. Two IDTs are designed such that two counter-propagating SAW modes are focused at a central point\cite{maity_coherent_2020}, as shown in Fig. \ref{fig1}(a). We use an SiV located near this point to maximize its coupling efficiency to both modes, one of which is used for the continuous dressing field and the other for the pulsed probing field. (More details about the device can be found in supplementary section \ref{ssec:saw_device}). The sample is placed in a dilution refrigerator with a base temperature of $\sim$100 mK to suppress thermal phonons resonant with the SiV’s orbital transition of $\sim$50 GHz. These thermal phonons are known to limit the spin coherence when the SiV is at temperatures above $\sim$1K\cite{jahnke_electronphonon_2015,sukachev_silicon-vacancy_2017}. A superconducting vector magnet applies an external magnetic field to Zeeman-split the spin levels into resonance with the IDT’s frequency. We set the magnetic field perpendicular to the symmetry axis of the SiV in order to maximize the spin-phonon coupling\cite{meesala_strain_2018} and obtain more effective noise suppression (see supplementary section \ref{ssec:coherence_analysis} for more information).

A simplified electronic level structure of a single SiV is shown in Fig. \ref{fig1}(b). In the absence of driving, the spin states are bare levels $\ket{0}$ and $\ket{1}$ with a Zeeman splitting tunable by an external magnetic field. We apply a continuous acoustic field resonant with this splitting and form new dressed eigenstates $\ket{\pm}=\frac{1}{\sqrt{2}} (\ket{0} \pm \ket{1})$. In contrast to previous work\cite{miao_universal_2020,laucht_dressed_2017}, we have resonant optical access to each dressed state, shown in the spectra in Fig. \ref{fig1}(c). Both excited states $\ket{e_0}$ and $\ket{e_1}$ decay to the dressed states, so we can optically pump the system into $\ket{+}$ ($\ket{-}$) by applying a resonant laser to an optical transition corresponding to $\ket{-}$ ($\ket{+}$). This can be used for direct state initialization and readout, which alleviates the need for adiabatic ramp-up and ramp-down of the dressing field. This fast and simple technique will be particularly useful for future phononic quantum networks\cite{arrazola_toward_2024}. 

While the resonant dressing acoustic field with frequency $\omega$ is on, we can apply a weaker probing acoustic field with frequency $\omega_p$. When both fields are on, the system is described by the Hamiltonian:
\begin{equation}\label{eq1}
    H = \frac{\omega}{2}\sigma_z +  \Bigl[\Omega^D\sigma_x + \Lambda^D\sigma_z\Bigr]\cos(\omega t)  + \Bigl[\Omega^P\sigma_x + \Lambda^P \sigma_z\Bigr] \cos(\omega_p t)
\end{equation}
where $\sigma_x$ and $\sigma_z$ are the Pauli matrices. The dressing (probing) field can be decomposed into two polarization components, one that drives the bare states with Rabi frequency $\Omega^D$ ($\Omega^P$), and one that modulates the energy splitting of the bare states with amplitude $\Lambda^D$ ($\Lambda^P$). In the dressed state basis and under the rotating wave approximation (see supplementary section \ref{ssec:Hammie_theory} for details), the Hamiltonian transforms to:
\begin{equation}
\tilde H = \frac{\Omega^D}{2} \tilde\sigma_z + \Lambda^P\cos(\omega_p t) \tilde\sigma_x + \frac{\Omega^P}{2}\Bigl(\tilde\sigma_y\sin((\omega-\omega_p) t)   +\tilde\sigma_y\sin((\omega+\omega_p) t) \Bigr) \end{equation}
where $\tilde\sigma_x$, $\tilde\sigma_y$, and $\tilde\sigma_z$ are the Pauli matrices in the dressed state basis. Here we also eliminate the time-dependent $\tilde\sigma_z$ term because its effect is negligible in the limit where $\Omega^P \ll |\omega\pm\omega_p|$.
In the rotating frame, the dressed states are split in energy by $\Omega^D$. However, we observe that the dressed states can be resonantly driven by a probing field with frequency $|\omega_p|=\Omega^D$ or with frequency $|\omega_p|=\omega \pm\Omega^D$, and in this sense they have multiple different resonances in the lab frame (Fig. \ref{fig1}(b)). In this work we set the probing field frequency to be $\omega_p = \omega -\Omega^D$. After neglecting far-off resonant driving terms, the system Hamiltonian is further simplified to the canonical resonantly driven two-level system:
\begin{equation}\label{eq3}
\tilde H = \frac{\Omega^D}{2} \tilde\sigma_z + \frac{\Omega^P}{2} \sin(\Omega^D t) \tilde\sigma_y .
\end{equation} Resonant driving of dressed states has been demonstrated at the smaller splitting frequency $\Omega^D$ in past experimental demonstrations\cite{laucht_dressed_2017,kim_suppression_2025,xu_coherence-protected_2012}, but this derivation indicates that the dressed states can equivalently be driven at the larger splittings $\omega \pm\Omega^D$, which we experimentally confirm in the below results. While demonstrated here for a classical field, this feature also enables the coupling the spin to a near-resonant phonon mode. 

\begin{figure}
\centering
\includegraphics[width=0.7\textwidth]{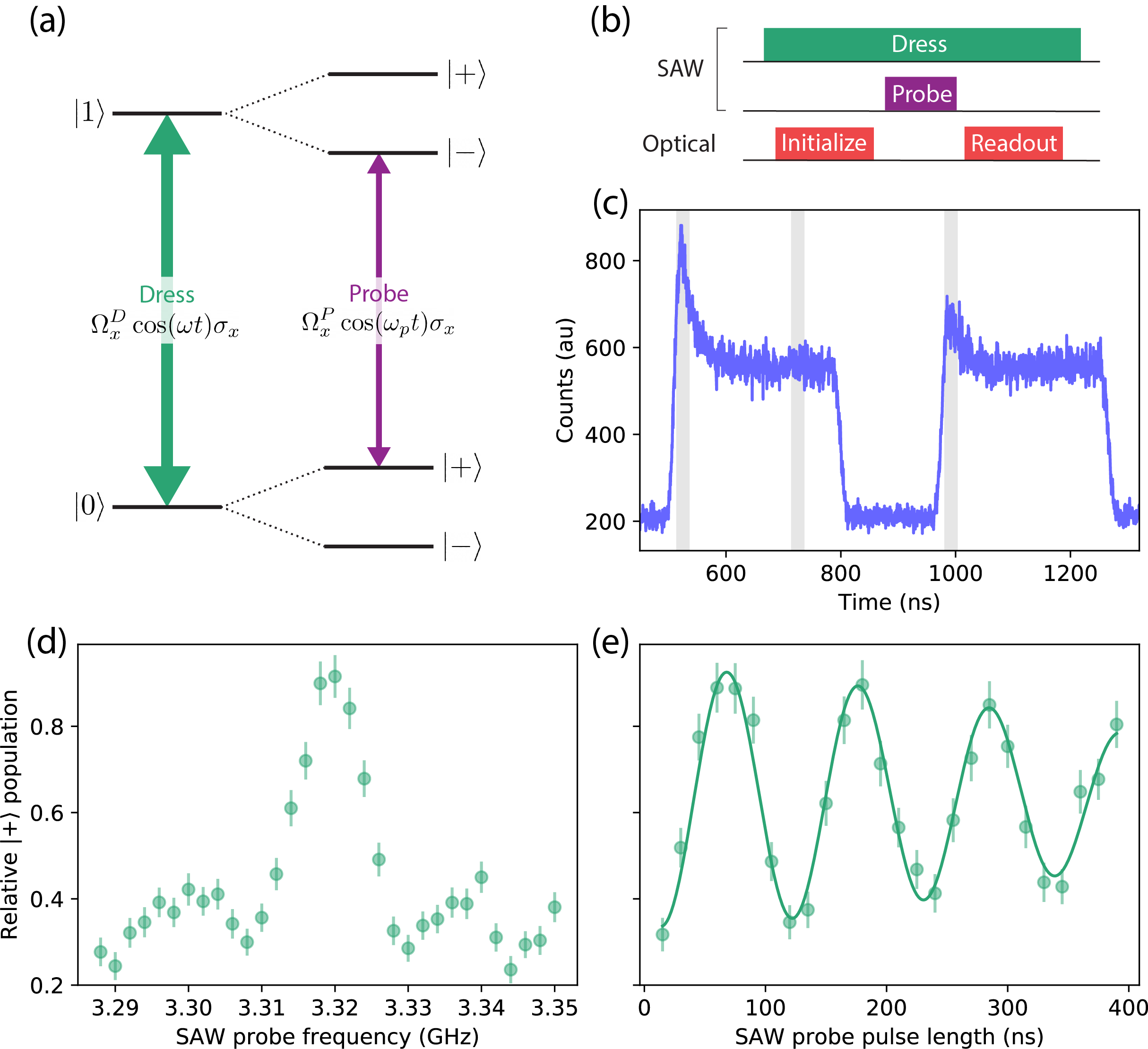}
\caption{Coherent control of a dressed spin qubit. \textbf{(a)} The level structure of the bare and dressed states. \textbf{(b)} The measurement sequence for an optically detected acoustic resonance (ODAR) and Rabi measurement. The dressing SAW is on throughout, and a probing SAW pulse rotates the state in between laser initialization and laser readout. \textbf{(c)} A histogram of photon counts from an optical measurement sequence. The decrease in counts after the rising edge of each pulse results from the optical pumping of the spin population. The gray shaded regions indicate the 20ns time windows over which the counts are integrated to calculate the relative state population. \textbf{(d)} An optically detected acoustic resonance (ODAR) measurement. The peak corresponds to the transition at frequency is $\omega -\Omega^D$ between dressed spin states $\ket{+}$ and $\ket{-}$. The probing SAW pulse has a length of 67 ns and a power of $\sim8\mu$W at the SiV location. \textbf{(e)} A resonantly driven Rabi oscillation between dressed states with a fitted Rabi frequency of $9.2\pm0.1$MHz. The probing SAW power is $\sim5\mu$W at the SiV location. For (c)-(e), the bare state resonance $\omega$ is 3.394GHz, and the dressing Rabi frequency $\Omega^D$ is 76MHz.}  \label{fig2}
\end{figure}

We characterize the dressed states with the pulse sequence shown in Fig. \ref{fig2}(b). An initialization laser pulse first optically pumps the population into $\ket{-}$, and after applying a SAW pulse with the probing field, we readout the state population via a second laser pulse. The relative population in $\ket{+}$ is calculated by integrating the counts in the 20ns window at the beginning of the readout pulse, adjusting for the background steady state laser counts, and normalizing to the counts in the 20ns window at the beginning of the initialization pulse (Fig. \ref{fig2}(c)). The initialization fidelity is limited by the optical linewidth, which is on the order of 100MHz, such that a laser resonant with one transition also partially drives the nearby transition. More information on the measurement method can be found in supplementary section \ref{ssec:histograms}.

The dressed state spin resonance at $\omega -\Omega^D$= 3.32 GHz is obtained via an optically detected acoustic resonance (ODAR) sequence (Fig. \ref{fig2}(d)), in which the frequency of the probing SAW pulse is swept. When the probing SAW is resonant with the splitting between the two dressed states, the population is transferred from $\ket{-}$ to $\ket{+}$ and we observe a peak. The smaller features of the ODAR off of the main peak are due to the frequency-dependent response of the IDT convolved with off-resonant driving effects. We also demonstrate Rabi oscillations between the dressed states by fixing the probing SAW frequency on resonance with the dressed state transition and varying the duration of the probing SAW pulse (Fig. \ref{fig2}(e)). The Rabi frequency of the dressed state is fitted to be $9.2\pm0.1$MHz. This Rabi rate is comparable to or greater than other demonstrated Rabi oscillations of dressed spin states in solid-state materials\cite{laucht_dressed_2017,xu_coherence-protected_2012,barfuss_strong_2015}. 

\begin{figure}
\centering
\includegraphics[width=0.7\textwidth]{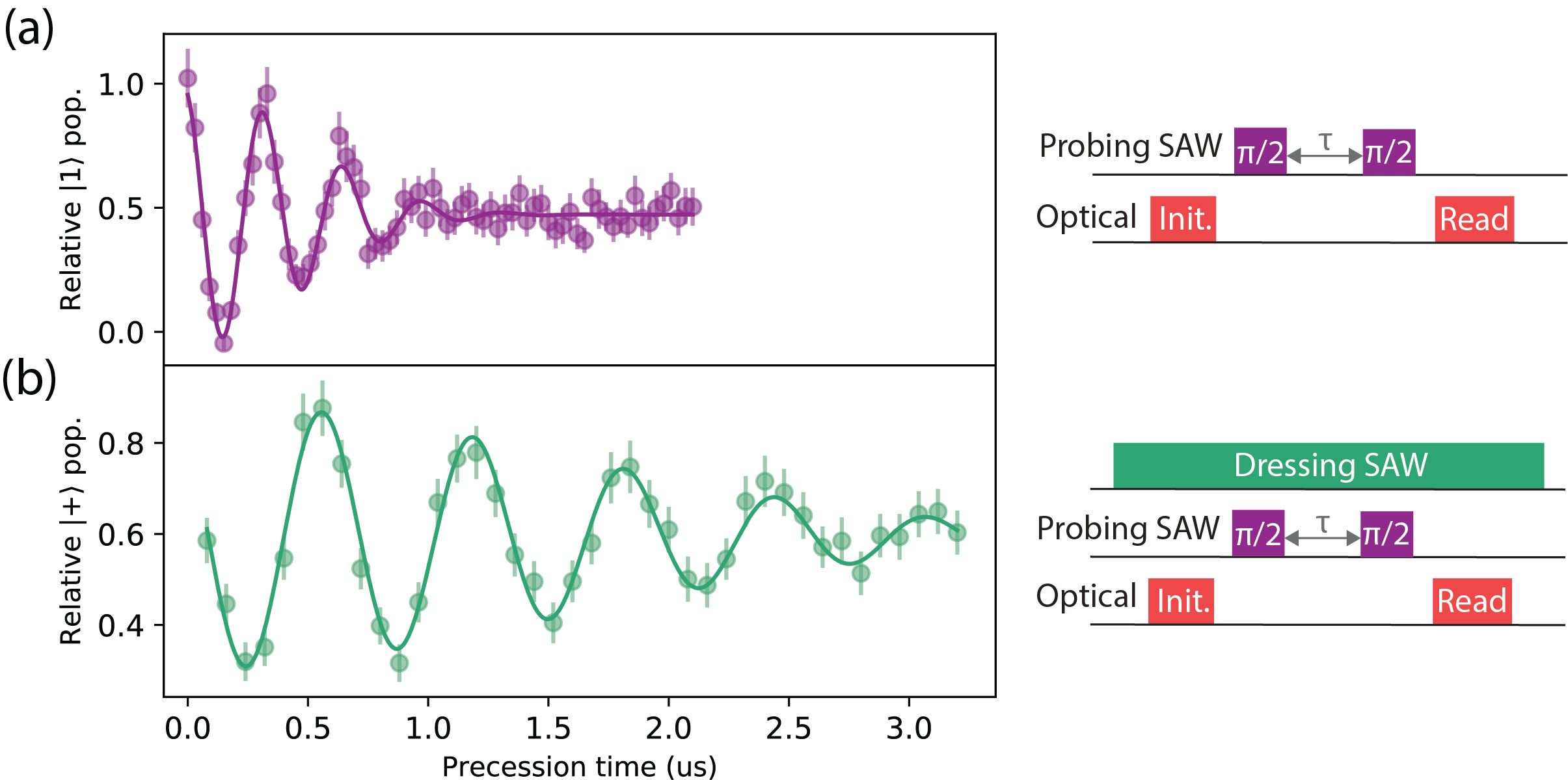}
\caption{Spin coherence extension of the dressed states. \textbf{(a)} Ramsey measurement of the bare states where the frequency of the $\pi/2$ pulses is intentionally detuned from $\omega$ by $\sim$3 MHz to increase the visibility of the decay envelope. The pulse sequence is shown on the right side of the figure. The fitted $T_2^*$ time is $680\pm40$ns. \textbf{(b)} Ramsey measurement applied to the dressed states with $\Omega^D=76$MHz and with the frequency of the $\pi/2$ pulses intentionally detuned from $\omega_p$ by $\sim$1.6 MHz. The fitted $T_2^*$ time is $2.2\pm0.2\mu$s. The detuning for this measurement is reduced so that fewer points can be measured while maintaining a clear signal.}  \label{fig3}
\end{figure}

In the bare state basis and at cryogenic temperatures, low-frequency magnetic field fluctuations from the environment limit the spin decoherence\cite{sukachev_silicon-vacancy_2017}. This noise contribution can be written as a term $\eta\sigma_z$ where $\eta$ is the noise amplitude. In the rotated dressed state basis, these same fluctuations will act as a $\tilde\sigma_x$ operator on the dressed states and in the off-resonant limit will scale as an AC Stark shift, with a quadratically suppressed noise amplitude of $\eta^2/|\Omega^D-\omega_n|$ where $\omega_n$ is the frequency of the noise. This reduced noise causes the dressed states to have a longer coherence time than the bare states, which we measure via a Ramsey sequence (Fig. \ref{fig3}). We apply two $\pi/2$ pulses and sweep the free precession time $\tau$ between them to obtain a Ramsey curve for the bare spin states and the dressed spin states (Figs. \ref{fig3}(a) and (b) respectively). We fit each curve to a sine wave damped by $\exp(-(\tau/T_2^*)^2)$ and extract a bare state $T_2^*$ time of $680\pm40$ns and a dressed state $T_2^*$ time of $2.2\pm0.2\mu$s. The external static magnetic field is the same for both measurements, and the temperature measured at the sample stage during the dressed state measurement is 250mK, compared to 45mK during the bare state measurement. This is strong evidence that the dressing field suppresses the noise susceptibility of the spin despite the additional heat it imparts to the sample.

It is crucial to limit this heating such that the SiV's temperature remains in the regime $<$1 K where the fast orbital phonon decoherence mechanism is suppressed \cite{jahnke_electronphonon_2015, sukachev_silicon-vacancy_2017}. 
At the same time, enough dressing power must be applied to create optically distinguishable dressed states and to extend the spin coherence time (see supplementary section \ref{ssec:coherence_analysis}). We are able to achieve both aims because of the high-efficiency of strain-based spin driving. Our SAW platform gives a Rabi rate scaling of $>$1GHz/$\sqrt{W}$ (with power at the input of the dilution refrigerator), enabling Rabi rates on the order of 100MHz with only modest waiting times to allow the sample to cool down. The experimental duty cycle for the data in Fig. \ref{fig3}(b) is 40\%. The best reported microwave driving efficiencies of the SiV spin are on the order of 10 MHz/$\sqrt{W}$ \cite{stas_robust_2022,sukachev_silicon-vacancy_2017}. At this scaling, applying a dressing field with a 100MHz Rabi frequency over a several-$\mu$s duration is experimentally infeasible.

\begin{figure}
\centering
\includegraphics[width=0.7\textwidth]{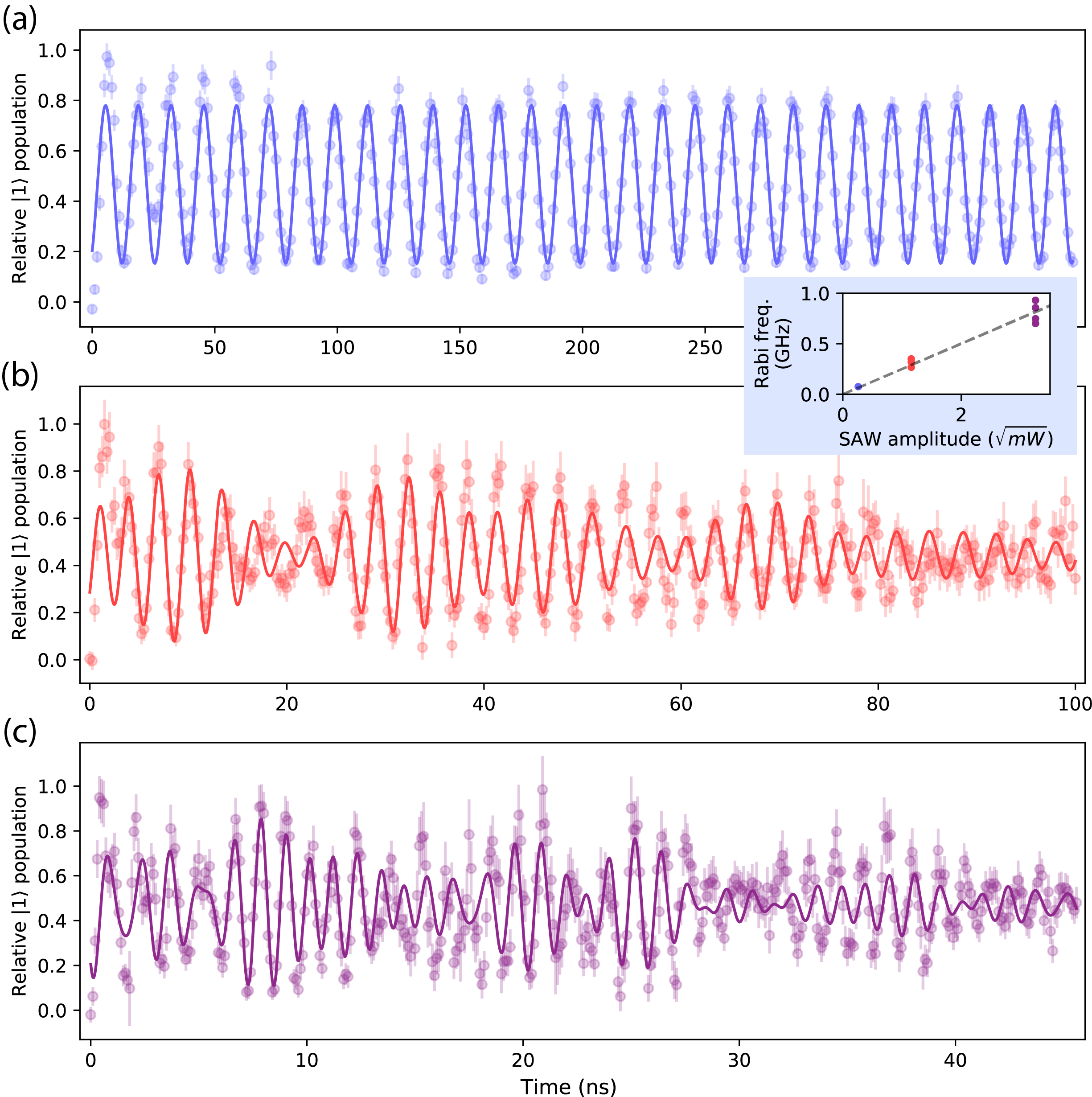}
\caption{Ultrafast coherent control of a bare spin state. The SAW powers are \textbf{(a)} 70uW, \textbf{(b)} 1mW, and \textbf{(c)} 10mW and lead to coherent oscillations with frequencies centered around \textbf{(a)} 76MHz, \textbf{(b)} 320MHz, and \textbf{(c)} 800MHz behavior. Increasingly complicated and multitone behavior is observed as the SAW power increases. The data is fit to sums of sine waves. \textbf{(Inset)} The frequencies of the fitted data are plotted against the SAW amplitude at the SiV location. The linear fit has a slope of 250 MHz/$\sqrt{mW}$.}  \label{fig4}
\end{figure}

In addition to demonstrating the extended coherence time of the dressed spin qubit, we probe the bare spin qubit’s behavior under strong mechanical driving. We observe ultrafast coherent dynamics on the order of 1 GHz (Fig. \ref{fig4}), the fastest ever Rabi frequency from a direct resonant drive of an SiV spin, comparable to the extremely fast all-optical control\cite{becker_all-optical_2018}. Multi-frequency behavior begins to appear in the Rabi measurement when the Rabi frequency is above 100 MHz. This beating behavior is not predicted by the assumed two-level model or by an off-resonant admixture of the 50
GHz orbital states. We also do not expect reflections of the SAW signal from the IDTs to be responsible for this effect since the amplitude of the reflected signal is an order of magnitude lower than the initial signal (see supplementary section \ref{ssec:saw_device}).
Other potential causes are: (1) The SAW pulse resulting from the IDT is not perfectly square and it deviates more from the ideal as the pulse gets shorter, limited by the bandwidth of the IDT. (2) The nonlinear phononic response of the AlN\cite{lu_non-degenerate_2021,schneider_frequency_2020} may cause higher-order driving effects.

\section{Discussion and Outlook}
In conclusion, we demonstrate control and readout of a dressed SiV spin, exploiting the high SiV strain susceptibility. The spin is protected by the dressing field such that we observe a more than threefold extension of coherence time. The dressed states have additional promising characteristics, such as a fast and easy tunability which can be achieved by sweeping the amplitude of the dressing field. This could allow quanta of information to be controllably exchanged between SiV spins and phonon modes at different frequencies. Additionally, a single dressing field could be applied to two SiVs with different bare spin frequencies such that their dressed spins are both resonant with a specific cavity or waveguide mode (see supplementary section \ref{ssec:proposal}). 

The extended coherence time of the dressed states may be limited by a weak admixture of the orbital states, which have a relatively short, phonon-induced lifetime and are coupled strongly though far off-resonantly by the dressing SAW (see Supplementary materials for more information). Future improvements to the coherence time may thus be realized with an SiV in a different strain environment or with other defects. For example, a static DC strain in the SiV’s environment can increase the orbital splitting\cite{meesala_strain_2018} and reduce potential decoherence effects from orbital states\cite{sohn_controlling_2018}. This must be balanced, however, with the reduced strain susceptibility in higher-strain SiVs\cite{meesala_strain_2018}.  The dressed state spin qubit protocol may also apply to the negatively charged tin vacancy (SnV) center, whose orbital splitting is 850 GHz in a low strain environment and can be extended to 1300 GHz with applied strain\cite{guo_microwave-based_2023}. The larger frequency difference between the spin splitting and the orbital splitting could improve the dressed coherence time extension in the SnV spin, though its SAW-spin susceptibility has yet to be experimentally demonstrated. Alternatively, this noise mechanism could be suppressed by engineering the surrounding phononic density of states to suppress the orbital state decay pathways. This has been demonstrated in the past to extend the ground-state orbital lifetime of the SiV\cite{kuruma_controlling_2025}.

The coherent dynamics measured in the bare states, limited by the power output of our microwave amplifier, set a new record for the SiV spin. Despite their multi-tone behavior, these fast dynamics show a path forward to ultrafast  control, which is generally useful for reducing gate times and could also pave the way for an alternative noise decoupling scheme which relies on fast decoupling pulses to be compatible with a spin-phonon cavity \cite{arrazola_engineering_2025}. Such strategies require the implementation of $\pi$ pulses that are at least two or three orders of magnitude shorter than $1/g$, where $g$ is the coupling strength between the spin and the mechanical cavity. For a spin-cavity coupling rate of $g$=300kHz as has been demonstrated for SiV spins in mechanical cavities\cite{joe_observation_2025}, this scheme requires pulse lengths in the range from a few tens of nanoseconds, which is already achievable with this system, to a few nanoseconds, which could be reached with only modest device improvements. For example, the acoustic driving efficiency can be improved by using lithium niobate as the piezoelectric material, which has higher electromechanical coupling efficiency\cite{xu_thin_nodate}. This would reduce the required number of IDT fingers\cite{aref_quantum_2016}, leading to a larger IDT bandwidth and a shorter ramping time of the SAW pulses. In addition, with relatively modest increases in the Rabi driving rate, this system could be used to explore the nonlinear behavior of a spin qubit in the regime where the rotating frame approximation breaks down\cite{fuchs_gigahertz_2009}.

\section{Methods}
The sample is mounted in a dilution refrigerator. The schematic of the measurement setup is shown in Fig. \ref{figS1}. We use a home-built confocal optical system to initialize and read out the SiV spin states.  The SiV optical transitions are driven by two tunable lasers around 737 nm (Toptica DL Pro). One laser serves as a repump laser (refer to supplementary section \ref{ssec:optical} for more information). Both lasers are gated by fiber-based acousto-optic modulators (AOM,  AA Optoelectronics MT200-R18-Fio-PM-J1-A-VSF) which create the optical pulses. An additional green laser at 520 nm is used for charge stabilization of the SiV and is gated with a pulse length of 1 ms and a 10\% duty cycle by a delay generator (DG, Stanford Research Systems DG645). Its peak power is $\sim$400$\mu$W. The green laser and the resonant lasers are combined and sent through an objective lens in the dilution refrigerator (Bluefors LD250). The objective lens is mounted on a nanopositioner stack (Attocube ANPxz101) to position the objective lens relative to the diamond sample. The fluorescence on the SiV phonon sideband is spectrally filtered out from the laser reflection and collected by an avalanche photodiode (APD, Perkin Elmer). The APD is also gated inversely to the green laser, such that no counts are collected when the green laser is on. This is to reduce noisy photon counts coming from the fluorescence caused by the green laser in addition to residual reflection.

An arbitrary wave generator (AWG, Tektronix AWG70001A) synthesizes the dressing field microwave signal (around 3.39 GHz) which then passes through an amplifier (Mini-circuits ZHL-1W-63-S+) with 37.5dB amplification. The amplified signal then passes through an isolator and a bandpass filter before being sent into the dilution refrigerator and transduced by an IDT into a SAW signal. The isolator has 1dB insertion loss and the bandpass (BP) filter has 2dB insertion loss, so the total amplification of the path is 34.5dB. When the dressing Rabi frequency $\Omega^D$ is 76 MHz, as in Figs. \ref{fig2} and \ref{fig3}, the peak power at the microwave input of the dilution refrigerator is 2.3mW. The probing field microwave signal is synthesized by a high power source (Hittite HMC-T2240) and the pulses are created by a microwave switch (Mini Circuit ZASWA2-50DR-FA+).

\section{Contributions}
The experiment was conceived and designed by E.C., Z.X, P.R., B.P., and M.L. The measurements were done by E.C., Z.X., E.M., H.W., and M.H. The modeling and theory were done by Z.W., E.C., Z.X., L.J., P.R., and B.P. The device was fabricated by S.M. and the dilution refrigerator setup was designed and built by G.J., M.H., and H.W. The manuscript was written by Z.X. and E.C. with contributions from all authors.

\section{Acknowledgments}
The authors would like to acknowledge Matthew Yeh, Guanhao Huang, Yan Qi Huan, Nicholas Achuthan, and Amirhassan Shams-Ansari for useful discussions. We are grateful to Iñigo Arrazola and the other authors of the theory paper \cite{arrazola_toward_2024} which inspired this project. We also give our thanks to Marie Wesson and the Yacoby group for the use of their wirebonding tool and to Chang Jin for his assistance with lasers. 

This research was supported by the National Science Foundation under grant number EEC-1941583, the Air Force Office of Scientific Research under award numbers FA9550-23-1-0333 and FA9550-23-1-0338, Amazon Web Services under award number A50791, and the Packard Foundation under award number 2020-71479. This research is also part of the Munich Quantum Valley, which is supported by the Bavarian state government with funds from the Hightech Agenda Bayern Plus. H. K. W. acknowledges financial support from the NSF graduate research fellowship program. E.M. is supported by the Draper Scholar fellowship. G. J. was supported in part by the Natural Sciences and Engineering Research Council of Canada (NSERC). B. P. acknowledges financial support from Q-NEXT, a U.S. Department of Energy Office of Science National Quantum Information Science Research Centers under award number DE-FOA-0002253. This work was performed in part at the Center for Nanoscale Systems (CNS), a member of the National Nanotechnology Infrastructure Network (NNIN), which is supported by the National Science Foundation award ECS-0335765. CNS is part of Harvard University.

\bibliography{bib_v3}

\pagebreak
\widetext
\newpage

\section*{Supplementary Information}

\setcounter{section}{0}
\renewcommand{\thesection}{S\arabic{section}}
\setcounter{figure}{0}
\renewcommand{\thefigure}{S\arabic{figure}}
\setcounter{table}{0}
\renewcommand{\thetable}{S\arabic{table}}
\setcounter{equation}{0}
\renewcommand{\theequation}{S\arabic{equation}}

\section{A two-level system under dressing and probing drives}\label{ssec:Hammie_theory}
The spin coupled to the dressing field and probing field is described by the Hamiltonian in Eq. \ref{eq1}, reproduced here for the reader's convenience:
$$H = \frac{\omega}{2}\sigma_z +  \Bigl[\Omega^D\sigma_x + \Lambda^D\sigma_z\Bigr]\cos(\omega t)  + \Bigl[\Omega^P\sigma_x + \Lambda^P \sigma_z\Bigr] \cos(\omega_p t).$$
In general the time-dependent $\sigma_z$ term in the dressing drive creates additional Floquet quasienergies spaced by integer multiples of $\omega$. In our parameter regime, however, this effect will be negligible because the Floquet amplitude scales with $\Lambda^D/\omega \ll 1$ and because the additional quasienergies are far off-resonant from the other resonant driving terms, so for analytical simplicity we neglect the $\Lambda^D\sigma_z$ term. Then the full system dynamics are:
\begin{eqnarray}
    i\dot{c}_0 &=& -\frac{\Lambda^P}{2}(e^{i\omega_p t} + e^{-i\omega_p t}) c_0 + \Bigl[\frac{\Omega^D}{2}(e^{i\omega t} + e^{-i\omega t}) + \frac{\Omega^P}{2}(e^{i\omega_p t} + e^{-i\omega_p t})\Bigr] c_1 \nonumber \\
    i\dot{c}_1 &=& \omega c_1 + \frac{\Lambda^P}{2}(e^{i\omega_p t} + e^{-i\omega_p t}) c_1 + \Bigl[\frac{\Omega^D}{2}(e^{i\omega t} + e^{-i\omega t}) + \frac{\Omega^P}{2}(e^{i\omega_p t} + e^{-i\omega_p t})\Bigr] c_0. \label{eqS1}
\end{eqnarray} 
where for mathematical ease we add to the diagonal terms an offset of $\omega/2$. 

We take the following transformation to eliminate the static bare state energy separation and to change to the dressed basis:
\begin{eqnarray}
c_0 &=& \frac{1}{\sqrt{2}}(c_+ - c_-)  \nonumber\\
c_1 &=& \frac{e^{-i\omega t}}{\sqrt{2}}(c_+ + c_-).  \label{eqS2}
\end{eqnarray}
After dropping the terms rotating at $2i\omega t$, we obtain these dynamics in the dressed basis:

\begin{eqnarray}
i\dot{c}_+ = \frac{\Omega^D}{2} c_+ + \frac{\Lambda^P}{2}(e^{i\omega_p t} + e^{-i\omega_p t})c_- &+& \frac{\Omega^P}{4}(e^{-i\Delta_- t} + e^{i\Delta_- t} + e^{-i\Delta_+ t} + e^{i\Delta_+ t}) c_+ \nonumber \\
&-& \frac{\Omega^P}{4}(-e^{-i\Delta_- t} + e^{i\Delta_- t} - e^{-i\Delta_+ t} + e^{i\Delta_+ t}) c_- \nonumber \\
i\dot{c}_- =  - \frac{\Omega^D}{2} c_- + \frac{\Lambda^P}{2}(e^{i\omega_p t} + e^{-i\omega_p t})c_+ &-& \frac{\Omega^P}{4}(e^{-i\Delta_- t} + e^{i\Delta_- t} + e^{-i\Delta_+ t} + e^{i\Delta_+ t}) c_-  \nonumber \\
&+& \frac{\Omega^P}{4}(-e^{-i\Delta_- t} + e^{i\Delta_- t} - e^{-i\Delta_+ t} + e^{i\Delta_+ t}) c_+
\end{eqnarray}
where $\Delta_\pm = \omega\pm\omega_p$. These dynamics correspond to the following Hamiltonian:
\begin{equation}
    \tilde H = \frac{\Omega^D}{2} \tilde\sigma_z + \Lambda^P\cos(\omega_p t) \tilde\sigma_x + \frac{\Omega^P}{2}\Bigl(\cos(\Delta_- t) \tilde\sigma_z  + \sin(\Delta_- t) \tilde\sigma_y + \cos(\Delta_+ t) \tilde\sigma_z  + \sin(\Delta_+ t) \tilde\sigma_y\Bigr)
\end{equation}
where $\tilde\sigma_z$, $\tilde\sigma_y$ are the Pauli matrices in the dressed basis and we observe that the system can be resonantly driven under various frequency conditions. If $|\omega_p|=\Omega^D$, the probing field will drive the dressed states with Rabi frequency $\Lambda^P$. If $|\omega_p|=\omega\pm\Omega^D$, the probing field will drive the dressed states with Rabi frequency $\frac{1}{2}\Omega^P$. In our experiment we chose $\omega_p=\omega-\Omega^D$ (i.e. we chose $\Delta_- = \Omega^D$), so we can neglect the driving terms $\Lambda^P\cos(\omega_p t) \tilde\sigma_x$ and $\frac{1}{2}\Omega^P \sin(\Delta_+ t) \tilde\sigma_y$ because they will be far off-resonant. We also operate in the low probing power limit where $\Omega^P \ll \Delta_\pm$, so we can neglect the time-dependent $\tilde\sigma_z$ terms, whose effect will scale with $\Delta_\pm/\Omega^P$. We are left with the simplified Hamiltonian
$$\tilde H = \frac{\Omega^D}{2} \tilde\sigma_z + \frac{\Omega^P}{2} \sin(\Omega^D t) \tilde\sigma_y$$
which is shown in Eq. \ref{eq3}.

We see that the dressed states and the bare states are coupled by the same SAW field polarization component. This deviates from the standard intuition that two states can only be coupled by a field that is orthogonal to their quantization axis. The quantization axis of the dressed states is defined by the $\sigma_x$ component of the dressing field, and yet the dressed states can also be coupled by the $\sigma_x$ component of the probing field. We can interpret this phenomenon as a result of the interference between the probing and dressing fields. The $\sigma_x$ components of the two fields interfere such that the probing field effectively modulates the phase and amplitude of the dressing field and in this way produces an effect equivalent to a coherent coupling.

\section{Measurement setup}\label{ssec:setup}
\begin{figure}
\centering
\includegraphics[width=0.7\textwidth]{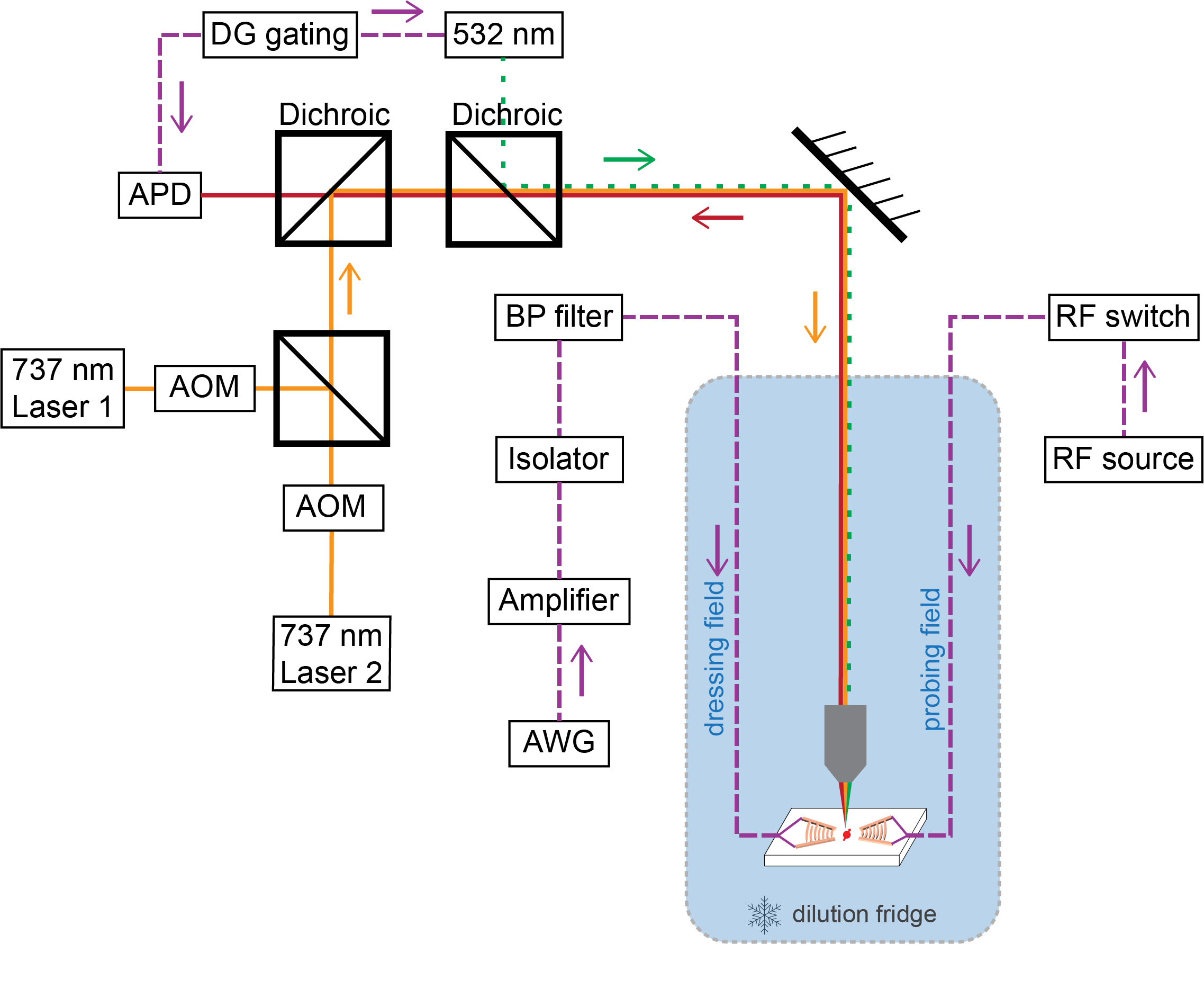}
\caption{Schematic of measurement setup. BP filter: Bandpass filter, AOM: Acousto-Optic Modulator, APD: Avalanche Photodetector, AWG: Arbitrary waveform generator, DG: Delay generator.}  \label{figS1}
\end{figure}

The measurement setup is shown in Fig. \ref{figS1}. Further information is located in the Methods section of the main text.

\section{Surface acoustic wave device}\label{ssec:saw_device}
\begin{figure}
\centering
\includegraphics[width=0.7\textwidth]{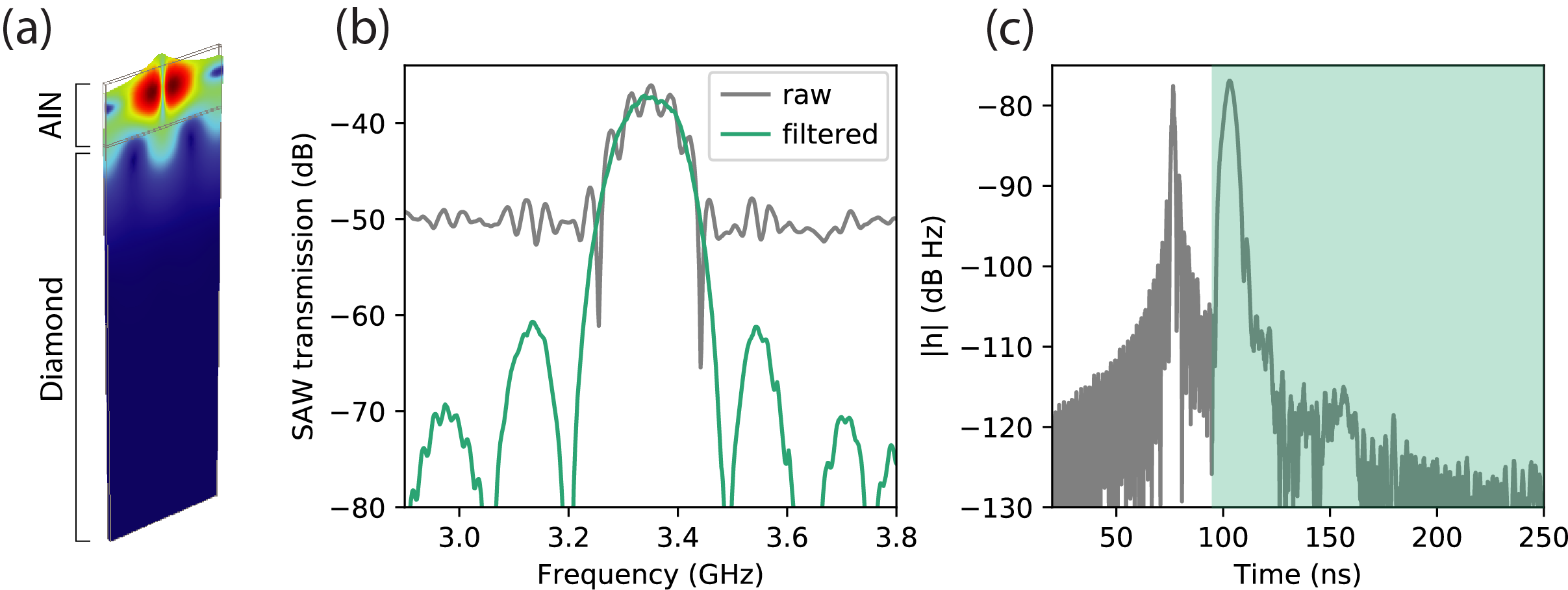}
\caption{SAW device characterization \textbf{(a)} Simulated mode shape of the SAW. The evanescent tail of the strain profile extends into the diamond and couples to the SiV. \textbf{(b)} Measurement of the microwave transmission S21 between two IDTs at $\sim$400mK. The bandwidth is 100MHz, and the loss is -18dB the dilution refrigerator to the IDTs is -18dB. \textbf{(c)} SAW transmission found via a fast fourier transform of (a). The fast-arriving sharp pulse is due to microwave cross-talk. The wider pulse arriving around 110ns is the zero-th order SAW propagation. The device parameters indicate that the first-order reflected pulse arrives around 155ns.}  \label{figS2}
\end{figure}

The SAW device is the same as used in a previous work\cite{maity_coherent_2020}. The wavelength of the SAW is 3 $\mu$m and the acoustic mode is confined to about one acoustic wavelength in the vertical direction, as shown in Fig. \ref{figS2}(a). A 1.4 $\mu$m layer of aluminum nitride (AlN) is sputtered onto a bulk electronic grade diamond to enable piezoelectrical tranduction of the SAW modes. The IDTs are designed to launch a beam that is Gaussian in the plane of the chip's surface in order to focus the acoustic energy to a waist of about one wavelength. The SiVs are generated $\sim$100 nm below the AlN-diamond interface by implantation of silicon ions and subsequent annealing. The implantation density is set such that there are only a few SiVs in an optical spot, and they are spectrally distinguishable such that they can be individually addressed with a resonant laser. 

The combined bandwidth of the two IDTs is $\sim$100MHz, as shown in the S-parameter characterization in Fig. \ref{figS2}(b). The total transmission loss from the whole SAW measurement path (measured at the dilution refrigerator input and output ports) is -37dB at the peak frequency of 3.35GHz. The loss comes from cable and wirebonding loss, device impedance mismatch, material loss, and SAW scattering into bulk modes. The estimated SAW powers at the SiV are determined from a calibration based on the Rabi frequency of the bare states at different frequencies and taking into account the slightly different responses of the two IDTs. The control pulses for bare state measurements and the continuous dressing fields applied for dressed state measurements are generated via the more efficient IDT, and the control pulses for dressed state measurements are generated via the other IDT.

The fast Fourier transform of the SAW transmission is shown in Fig. \ref{figS2}(c).  The first sharp pulse arriving at t$\sim$77 ns is due to microwave cross-talk. We can filter out the microwave crosstalk by selecting only the late-arriving part of the signal, highlighted in green, and taking the reverse Fourier transform to obtain a filtered frequency domain signal (Fig. \ref{figS2}(b)) corresponding only to the SAW frequency response.

The first SAW pulse arrives around t=103 ns so the propagation time of the SAW between IDTs is 26 ns. The IDT separation is d$\sim$230 $\mu$m, giving a SAW group velocity of 8.8 km/s. The first reflected SAW pulse arrives at t$\sim$155 ns and is attenuated by about 40dB from the initial SAW pulse. Because this loss is due to a full round trip, and assuming symmetric propagation through the device, this corresponds to $\sim$20dB attenuation at the SiV location between the initial pulse and the first reflected pulse.

\section{Optical measurement details}\label{ssec:optical}
\subsection{Optical spectra}\label{ssec:spectra}
We measure the optical spectra with continuous-wave two-laser photoluminescence excitation. We here refer to the two lasers as the repump laser and the initialization laser, named for the role they play in the pulsed measurement described in more detail in section \ref{ssec:histograms}. The repump laser is first parked at a fixed frequency resonant with the transition between the dressed state $\ket{-}$ and excited state $\ket{e_0}$. This laser continuously optically pumps the population out of $\ket{-}$ and into $\ket{+}$. Then as the frequency of the initialization laser is swept, the steady-state fluorescence signal features peaks that correspond to the transitions between $\ket{+}$ and the excited states (Fig. \ref{fig1}(c)). Next we park the repump laser on the transition between the dressed state $\ket{+}$ and excited state $\ket{e_0}$ and repeat the measurement in order to see the spectra corresponding to the state $\ket{-}$. We observe a splitting between the dressed states of  $\Omega^D\sim150$MHz, which we validate with an ODAR measurement similar to the one in (Fig. \ref{fig2}(d)). The dips in the spectra at the peaks corresponding to the  $\ket{-}$-$\ket{e_0}$ and $\ket{+}$-$\ket{e_0}$ transitions stem from coherent population trapping, which appears when the optical transitions excited by the two lasers share a single excited state. The repump and initialization laser power are each 1uW.

\subsection{Pulsed time-domain measurements}\label{ssec:histograms}
The laser power used in Fig. \ref{fig2}(c) is 400nW. Each laser pulse is 300 ns long. Before beginning each measurement sequence, we apply a repump laser pulse to transfer the spin population into $\ket{+}$. This allows the integrated counts during the sharp peak at the beginning of the initialization pulse to serve as a good normalization, because they correspond to all of the spin population (up to the initialization fidelity) in $\ket{+}$. The initialization fidelity is limited by the optical linewidth, which is on the order of 100MHz, so for dressed state splittings near that value a laser resonant with one transition also partially drives the nearby transition. 

\section{Optically detected acoustic resonance (ODAR) of bare states}\label{ssec:bareODAR}
\begin{figure}
\centering
\includegraphics[width=0.7\textwidth]{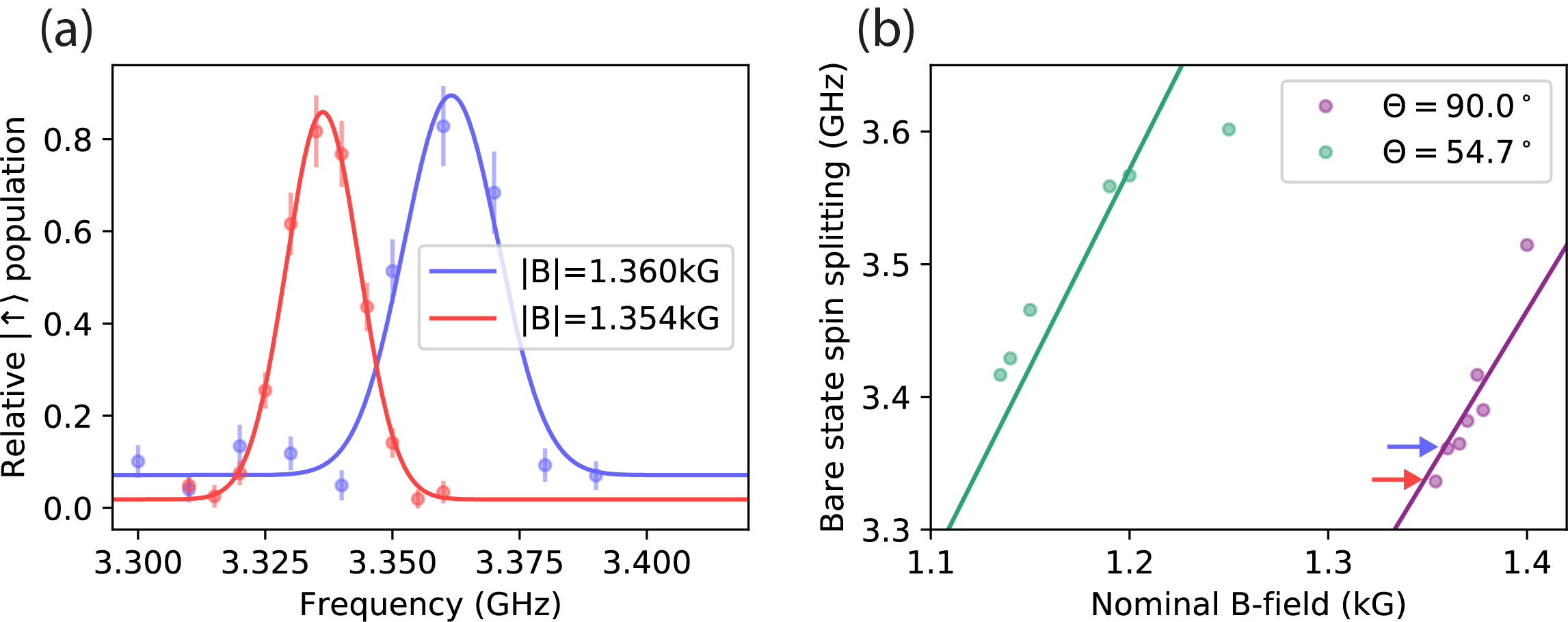}
\caption{\textbf{(a)} Two examples of bare state ODAR measurements used to find the bare state splittings, with the corresponding points indicated in (b). The red (blue) curve has a nominal magnetic field of 1.354 (1.360) kG orthogonal to the SiV symmetry axis and a fitted center frequency of 3.336 (3.362) GHz. The red (blue) curve’s points were measured with a pulse length of 50 (35) ns and SAW power of $\sim$0.7 ($\sim$3)  uW at the SiV. \textbf{(b)} The Zeeman tuning of the bare spin states with an externally applied magnetic field at an angle of 54.7° (90.0°) with respect to the symmetry axis of the SiV is plotted in green (purple), with a fitted slope of 3.0 (2.5) GHz/kG. The points are larger than the bare state splitting error bars.}  \label{figS3}
\end{figure}
To characterize the bare spin resonance, we employ an ODAR sequence on the undressed spin states. Two example ODAR traces are shown in Fig. \ref{figS3}(a). A Gaussian function is used to fit the resonance peak. In the example shown, the red and blue curves correspond to nominal external magnetic fields of 1.354 kG and 1.360 kG, respectively, both oriented orthogonally to the SiV symmetry axis. The fitted center frequencies are 3.336 GHz and 3.362 GHz, respectively. The extracted resonance frequencies as a function of the magnetic field are summarized in Fig. \ref{figS3}(b), where a clear Zeeman shift is observed. As expected, the slope of the Zeeman tuning varies depending on the angle of the external magnetic field. The observed deviations from a perfect linear dependence are attributed to magnetic field hysteresis and systematic errors in the superconducting magnet's current calibration, which we use to estimate the magnetic field at the SiV location. Our measurements are constrained by the device's SAW bandwidth, so we are only able to resolve spin resonances that lie within the finite acoustic frequency range supported by the IDT. This limitation restricts the total magnetic field range over which ODAR measurements can be performed.

\section{Coherence time dependence on dressing SAW amplitude and B-field}\label{ssec:coherence_analysis}
\begin{figure}
\centering
\includegraphics[width=0.7\textwidth]{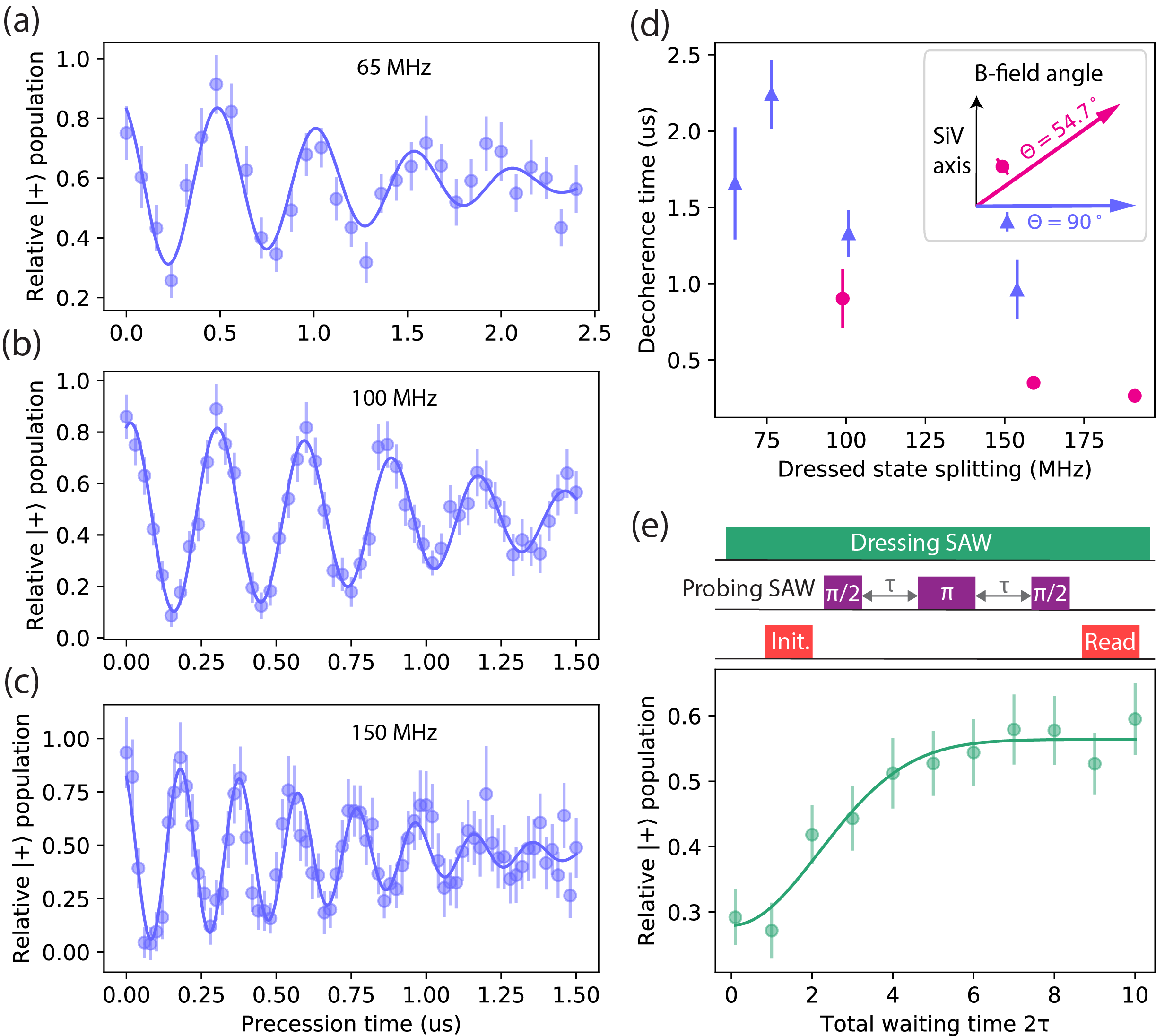}
\caption{\textbf{(a-c)} Ramsey measurements at dressed state splittings of 65, 100, and 150 MHz with a magnetic field perpendicular to the SiV symmetry axis. \textbf{(d)} Dressed state decoherence times plotted in blue (pink) are measured via Ramsey with magnetic field of 1.4 (1.2) kG at angle 90 (54.7)° with respect to the SiV. The most extended coherence time measured occurs at 75 MHz and $\theta$=90° (Fig. \ref{fig3}(b)), and this could be limited by thermal phonons, upper orbital states coupling, and magnetic field noise decoupling. \textbf{(e)} A spin echo measurement, with an additional refocusing $\pi$ pulse in between the $\pi/2$ pulses. The measurement is taken with a perpendicular magnetic field and a dressed state splitting of 76MHz, and has a coherence time $T_2$ of $3.1\pm0.8 \mu$s, similar to the coherence time  $T_2^*$ without refocusing pulses. This indicates that the dominant noise source fluctuates on timescales shorter than about a $\mu$s or is due to a state decay.
}  \label{figS4}
\end{figure}

We take a Ramsey measurement at various values of $\Omega^D$ (as shown in Fig. \ref{figS4}(a)-(c)) and observe that the coherence time is strongly dependent on the dressed state splitting (Fig. \ref{figS4}(d)). We consider three different factors that may limit the coherence time: (1) magnetic field noise decoupling, (2) thermal phonons, and (3) upper orbital states coupling. When $\Omega^D$ is too low, we expect that the noise suppression effect created by the dressed states is reduced, since the amplitude of low-frequency noise will scale with $1/\Omega^D$ (see the discussion about coherence time in the main text). The dephasing at low dressed state splittings could also be due to thermal phonons, which have a higher thermal occupation at lower frequencies. We could not measure dressed states with a splitting smaller than $\sim$60MHz because the optical transitions corresponding to different dressed states became indistinguishable.

As the dressing field amplitude increases, we can no longer neglect the upper orbital branch of the SiVs ground state, separated by $\sim$50GHz from the lower orbital branch. In addition to coupling the spin states, the dressing field couples the lower branch to the upper branch and the dissipation of the orbital states becomes relevant to the dressed state dissipation. In the off-resonant limit, the upper orbital states will have approximately $\Omega^2_{orb}/\Delta^2_{orb}$ of the state population, where $\Omega_{orb}$ is the orbital Rabi frequency and $\Delta_{orb}$ is the detuning between the dressing field and the orbital resonance, so we can estimate that the dressed state decoherence resulting from the upper orbital state dissipation $\gamma_{orb}$ will scale with $\gamma_{orb} \Omega^2_{orb}/\Delta^2_{orb}$. The orbital dissipation is highly temperature dependent and thus likely varies throughout the measurement sequence, but we can estimate the orbital decay rate as tens of MHz\cite{kuruma_controlling_2025,sohn_controlling_2018} and in this way get a minimum value for the orbital dissipation, which may be larger depending on the value of the orbital pure dephasing rate. The orbital states’ strain susceptibility is a few tens of times higher than that of the spin states\cite{meesala_strain_2018}, so for a spin Rabi frequency of $\Omega^D\sim$100MHz, the orbital Rabi frequency will be on the order of a few GHz. Therefore, the orbital state dissipation will limit the dressed state coherence time to a few microseconds, with the coherence time decreasing with larger dressing field amplitudes. 

We tentatively ascribe the coherence time’s magnetic field dependence to the fact that the spin-strain susceptibility changes with the orientation of the magnetic field. When the magnetic field is orthogonal to the SiV’s axis, the spin strain susceptibility is larger and less SAW power is needed to reach a given dressed state splitting. On the other hand, the orbital strain susceptibility does not have a strong dependence on magnetic field\cite{meesala_strain_2018}. Thus, for example, the measurement done at $\theta$=54.7° and dressed state splitting of 100 MHz and the measurement done at $\theta$=90° degrees and dressed state splitting of 160 MHz both have dressing SAW powers of $\sim$270$\mu$W, and correspondingly have similar coherence times.

We additionally characterize the limitations on dressed state coherence by applying a refocusing pulse between $\pi/2$ pulses under the parameters that give the longest Ramsey coherence time. We observe a negligible change in the coherence, indicating that the noise limiting the dressed states is not refocusable.

\section{Power handling}

For the 76 MHz dressing field strength used in Figs. \ref{fig2} and \ref{fig3}, we used a probe frequency at 3.318GHz and a dressing field frequency at 3.3944GHz, both of which are within the efficient frequency range of the IDTs (see section \ref{ssec:saw_device}). The parameters of the dressed Ramsey measurements shown in section \ref{ssec:coherence_analysis} are given in Table \ref{tabs1}. We determine the splitting between the bare state resonance and the dressed state resonance to obtain a precise value for the Rabi frequency of the dressing field. After each measurement sequence, we turn off the dressing field to allow the sample to cool down. The duty cycles for the measurements are chosen based on the total measurement power and the desired average temperature. For example, in the dressed Ramsey measurement in Fig \ref{fig3}(b), the dressing field is applied for a time varying from 1.262 to 4.283 $\mu$s and the repetition period varies between 3.2 and 11.0 $\mu$s, depending on the precession time of the Ramsey pulse sequences. For that data, we repeat each measurement sequence 6.2x$10^8$ times to suppress the uncertainty in the measurement.

\begin{table}[h]
\caption{Parameters for dressed Ramsey measurements}\label{tabs1}
\begin{tabular}{@{}cccc@{}}
\toprule
RF power at the input & Rabi frequency of  & Dressing field duty & Sample stage\\
to the fridge (mW) & dressing field (MHz) & cycle (\%) & temperature (mK) \\
1.4 & 65 & 80 & 280 \\
2.3 & 76 & 40 & 250 \\
4.1 & 100 & 20 & 240 \\
9.1 & 150 & 10 & 260
 \\
\botrule
\end{tabular}
\end{table}

\section{Optical spectra of the dressed states}\label{ssec:broad_spectra}
\begin{figure}
\centering
\includegraphics[width=0.7\textwidth]{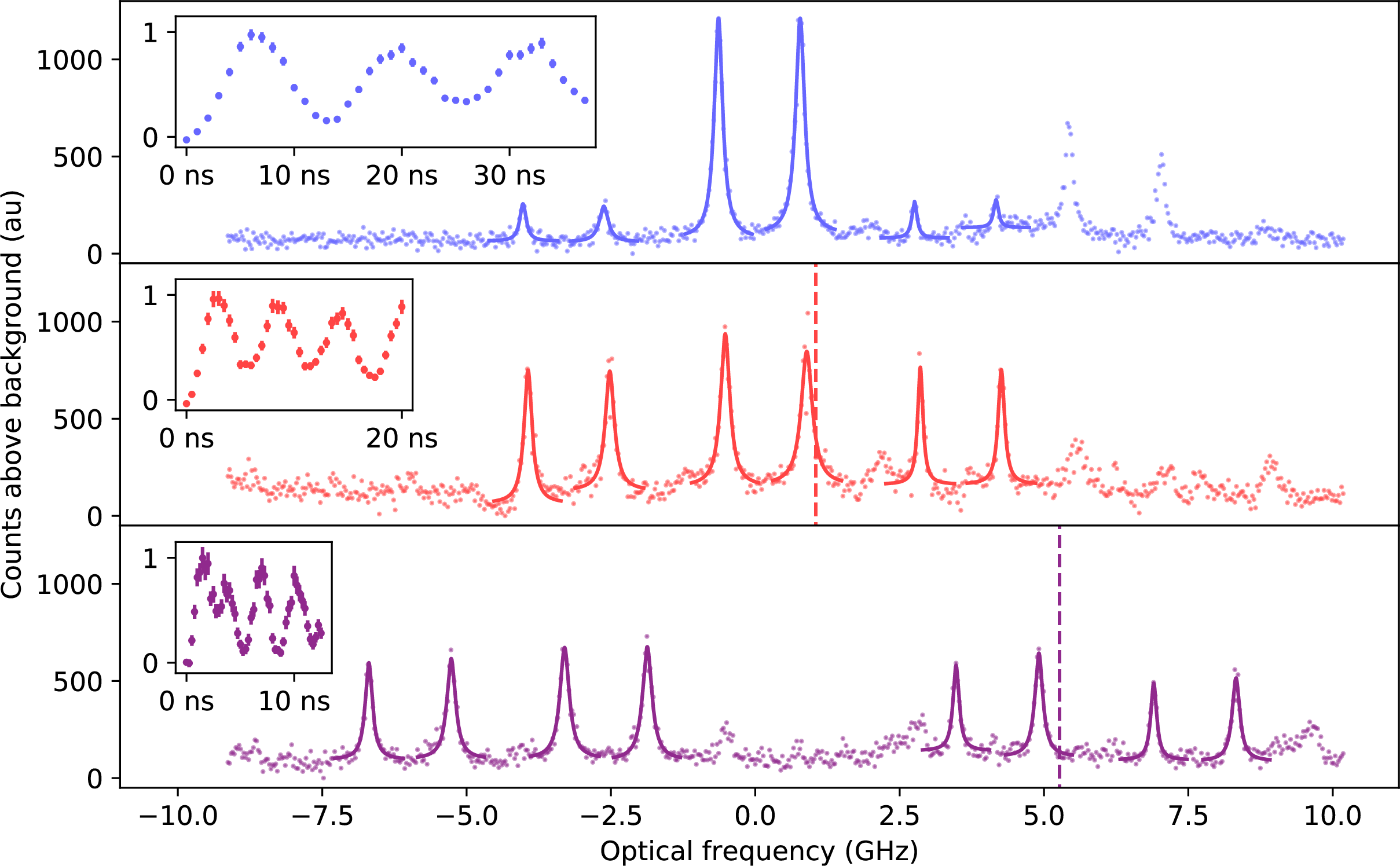}
\caption{Optical spectra at dressed state splittings of 76 (blue), 160 (red), and 320 (purple) MHz. The 76MHz spectrum is taken with a single laser because the dressed state energies are close enough that with 1uW of optical power a single laser continuously repumps the system between the two spin states. The 150 and 300MHz spectra are taken with a stationary repump laser on an optical transition corresponding to the $\ket{-}$ dressed state, so the peaks seen correspond to the $\ket{+}$ dressed state. The repump laser’s frequency is indicated with a vertical dashed line. All laser powers are 1uW. A Lorentzian fit to the data’s peaks is shown. We do not fit the peaks that correspond to other nearby SiVs in the sample. The bare state Rabi measurements taken at the corresponding SAW powers are shown in the inset and agree with the splitting measured in the optical spectra between the peaks corresponding to the $\ket{+}$ dressed state and those corresponding to the $\ket{-}$ dressed state (not shown). All data is taken with the magnetic field perpendicular to the SiV symmetry axis.}  \label{figS5}
\end{figure}

Along with the spin-strain response, the strain-induced modulation of the optical and orbital states cause shifts in the SiV’s optical spectrum. To investigate this strain-induced optical response, we measure photoluminescence spectra at dressed state splittings of 76 MHz, 160 MHz, and 320 MHz, as shown in Fig. \ref{figS5}. The corresponding bare state Rabi measurements, taken using the same SAW powers used to generate the dressed states, are shown in the inset of Fig. \ref{figS5}. For the 76 MHz condition, the dressed spin states are close enough in energy that a single laser at an optical power of 1 $\mu$W is sufficient to continuously repump the population between the two dressed spin states, resulting in a spectrum that reflects both states. At the larger dressed state splittings of 150 MHz and 320 MHz, we employ a stationary repump laser resonant with an optical transition associated with the $\ket{-}$ dressed spin state, so the peaks seen correspond to the $\ket{+}$ dressed state. The position of the repump laser is marked by a vertical dashed line. All measurements are performed with laser powers of 1$\mu$W. Lorentzian fits are applied to the primary peaks, excluding those attributed to neighboring SiV centers in the sample. As the dressing power increases, the extraneous peaks are eliminated from the spectrum because they are not optically repumped at the correct frequency.

The dressing SAW modulates the SiV optical transitions, producing sidebands spaced by integer multiples of the modulation frequency $\omega=3.394$ GHz. In addition to these sidebands, we observe a global blue shift of the entire optical spectrum. At the highest applied SAW power of $\sim$1mW, this shift reaches approximately 940 MHz. We analyze the underlying physical mechanisms that may contribute to this observed spectral shift, including the AC Stark shift and the pyroelectric effect in the aluminum nitride.

\subsection{AC Stark shift}
Strain modulates both the optical and the orbital levels of the SiV with an amplitude of $\sim$1 PHz/strain\cite{meesala_strain_2018}, so an AC Stark shift could result from either. Given the SiV spin-strain susceptibility of $\sim$100 THz/strain\cite{meesala_strain_2018}, we expect that the strain amplitude at the SiV location due to the dressing SAW is approximately 3e-6 for the data taken at the highest dressing power. The optical AC Stark shift arising from this SAW field is $\Omega_{opt}^2/\Delta_{opt}$, which is on the order of 10s of kHz. and the orbital AC Stark shift is equivalently $\Omega_{orb}^2/\Delta_{orb}$, which is on the order of 100s of MHz.

\subsection{Pyroelectric effect}
We also consider whether the pyroelectric effect could cause the observed optical shift. We perform the measurements of 76, 160, and 320 MHz with a dressing field duty cycle of 0.5, 0.2, and 0.05 respectively to avoid excessive heating of the dilution refrigerator, but this causes a changing temperature over the course of the measurement. AlN has a pyroelectric constant of $p \sim 7 \mu C/(m^2K)$\cite{fuflyigin_pyroelectric_2000}, a piezoelectric constant of $d_{33}=5.1$ pm/V\cite{lueng_piezoelectric_2000}, and a dielectric constant of $\epsilon$=9.2$\epsilon_0$\cite{thorp_dielectric_1990}. The pyroelectric strain in the AlN will be on the order of $pd_{33}\Delta T/\epsilon  = 4e-7 \Delta T /K^{-1}$. The strain at the SiV location in the diamond will be less than that in the AlN, but this gives an estimate of the size of the pyroelectric optical shift on the order of 100s of MHz for a $\sim$1K shift in temperature.

\section{SiV tuning scheme using dressed states
}\label{ssec:proposal}
Here we briefly discuss a proposal to resonantly couple two SiVs by taking advantage of the tunable nature of dressed states. Let us consider a mechanical cavity with two SiVs located inside of it, in which both SiVs have a non-zero coupling to several of the cavity modes. In order to resonantly couple the spins to each other via a long-lived phonon, we need to bring both SiVs into resonance with a cavity mode at $\omega_C$, but due to the variation in their strain environments, the SiV spins will in general have different frequencies under the same externally applied magnetic field. 

We can write the bare state spin splitting of the SiVs as $\omega_i = \epsilon_i B$ for $i=1,2$. We use an auxiliary cavity mode at frequency $\omega_A$ for the resonant amplification of a dressing field. Different SiVs will also generally have different susceptibilities to the dressing field in the cavity mode, depending on the spatial location within the cavity and SiV orientation, so a dressing field amplitude $A$ will dress them with Rabi frequencies $\Omega_i = \eta_i A$. A given SiV with a bare state frequency of i will have a dressed state resonance at, for example, $\tilde\omega_i = \omega_A +\Delta_i/2 - \sqrt{(\Delta_i/2)^2+ \Omega_i^2}$ where $\Delta_i = \omega_i-\omega_A$. (In the limit of zero detuning, the dressed state simplifies to $\tilde\omega_i = \omega_A  - \Omega_i$ and we recover the resonance shown in Fig 1(b).) By choosing an appropriate magnetic field and dressing field amplitude, we can simultaneously fulfill the condition $\omega_i=\omega_C$ for $i=1,2$ such that each SiV’s dressed spin is on resonance with the cavity mode at $\omega_C$.

\end{document}